\documentclass[aps,prb,twocolumn,groupedaddress]{revtex4}

\usepackage{latexsym}
\usepackage{graphicx}

\begin{document}


\title{Negative four-terminal resistance as a probe of crossed Andreev reflection}


\author{D. Beckmann}
\email[e-mail address: ]{detlef.beckmann@int.fzk.de}
\affiliation{Forschungszentrum Karlsruhe, Institut f\"ur Nanotechnologie, P.O. Box 3640, D-76021 Karlsruhe, Germany}
\author{H. v. L\"ohneysen}
\affiliation{Forschungszentrum Karlsruhe, Institut f\"ur Festk\"orperphysik, P.O. Box 3640, D-76021 Karlsruhe, Germany,
and Physikalisches Institut, Universit\"at Karlsruhe, D-76128 Karlsruhe, Germany}


\date{\today}

\begin{abstract}
We report on the experimental investigation of electronic transport in 
superconductor-ferromagnet spin-valve structures. Our samples consist of
two ferromagnetic iron leads forming planar tunnel contacts to a superconducting
aluminum wire. At energies below the superconducting gap, we observe a negative
four-probe resistance that can be explained by crossed Andreev reflection.
\end{abstract}


\maketitle


Quantum mechanics predicts a non-local correlation of spatially separated
particles that is stronger than that allowed by classical theory. 
This correlation, known as entanglement, has been tested experimentally 
on pairs of photons \cite{aspect1982} and massive particles \cite{rowe2001}, 
see also \cite{genovese2005}. An intriguing question is whether a stream
of spatially separated, entangled electron pairs can be created and 
manipulated in a solid-state environment, with possible future applications 
in quantum information processing.
A promising implementation of solid-state entanglers are
heterostructures involving a superconductor, 
\cite{recher2001,lesovik2001} as in a 
superconductor the electrons come naturally in the form of
Cooper pairs, which are in an entangled singlet state. The
challenge is to create structures where the two electrons
of a Cooper pair can be spatially separated without destroying
entanglement. If two normal-metal contacts are attached to a
superconductor at a distance smaller than the coherence
length of the superconductor, it has been predicted that an 
electron injected into the
superconductor may be transmitted into the second contact
as a hole, thereby creating a Cooper pair in the superconductor,
and leaving behind two entangled holes in the 
two normal metal contacts \cite{byers1995,deutscher2000}. 
There is a competition between this process, called crossed Andreev
reflection (CAR), and elastic cotunneling (EC), where the incident electron
gets transmitted to the second contact via a virtual state in the 
superconductor. Quantitative understanding of the contributions of CAR and competing
processes like EC is a prerequisite to building a solid-state entangler based on
superconductor hybrid structures. 

Crossed Andreev reflection has been studied experimentally in 
superconductor/ferromagnet\cite{beckmann2004} and 
superconductor/normal-metal\cite{russo2005} structures. 
In a recent experiment\cite{beckmann2004} on transport properties of 
superconductor/ferromagnet spin-valve structures, we have observed spin-dependent
transport in the superconductor at energies much smaller than the superconducting
gap. The observed signal decayed on the length scale of the coherence length of the
superconductor, and could be explained by theoretical predictions
for the superposition of crossed Andreev reflection 
and elastic cotunneling. Here, we present an experimental study that 
allows us to discriminate the two processes, and also rule out sequential
tunneling or non-equilibrium effects as a possible cause of the observed signals.

\begin{figure}
\includegraphics[width=\columnwidth]{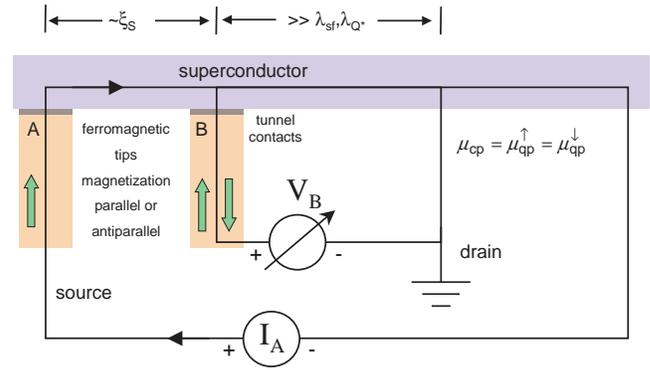}
\caption{\label{fig_scheme}(Color online) Schematic view of the experiment.
Two ferromagnetic leads form tunnel contacts to a superconducting
bar. Contact A is used to inject a current, and the voltage
at contact B is measured referenced to the drain voltage probe.
The magnetization alignment of the ferromagnets is either
parallel or antiparallel.}
\end{figure}

The schematic view of the experiment is shown in Fig. \ref{fig_scheme}. 
Two ferromagnetic wires A and B are attached via tunnel contacts to
a superconducting bar. Contact A (source) is used to inject a current into the 
superconductor, which then flows along the bar. The (Ohmic) drain contact, which lies 
inside the current path, is used as a reference of the chemical potential of the 
superconductor. Contact B is used to measure  
voltage relative to the drain contact. There are several possible mechanisms for the
observation of a finite voltage at contact B despite the fact that voltage is measured across
a superconductor: EC (CAR) emits an electron (hole) from the superconductor into contact 
B for an electron injected into contact A, thereby leading to a voltage that has 
equal (opposite) sign compared to the injector voltage. Both processes are elastic and
coherent, and involve virtual quasiparticle states in the superconductor, i.e. they
may occur at energies below the superconducting energy gap, and over the length scale of
the coherence length $\xi_\mathrm{S}$ of the superconductor. Note that for CAR, the chemical
potential of the detector contact B may lie outside the voltage window spanned by source
and drain contact, which has been predicted previously for a slightly different 
setup.\cite{jedema1999}

In addition to coherent subgap processes, an electron may tunnel into an allowed 
state in the superconductor and sequentially tunnel out at contact B.
Sequential tunneling, like elastic cotunneling, yields a 
voltage $V_\mathrm{B}$ of the same sign as $V_\mathrm{A}$.
Sequential tunneling requires an electron to have enough energy to overcome the
spectral gap, which can be supplied by the applied voltage, thermal
excitation or external noise. Alternatively, the gap may be smeared out by quasi-particle
life-time broadening, the inverse proximity effect in the contact regions, or the magnetic 
stray fields of the electrodes. Sequential tunneling may be either elastic and coherent,
or incoherent, generating a non-equilibrium charge \cite{clarke1972,tinkham1972} 
and spin \cite{johnson1994} accumulation in the 
superconductor. It is known that non-equilibrium quasiparticle populations in
superconductors relax only slowly due to electron-phonon scattering,\cite{chi1979} and 
accordingly the decay length of the non-equilibrium can be quite large.
From the data of our previous experiment,\cite{beckmann2004} we obtain the charge 
imbalance relaxation length $\lambda_\mathrm{Q^*}\approx 3~\mathrm{\mu m}$ and the 
normal-state spin diffusion length $\lambda_\mathrm{sf}\approx 1~\mathrm{\mu m}$.
A $10~\mathrm{\mu m}$ distance between source and drain contact was chosen,
to ensure that any non-equilibrium quasi-particle populations which may be injected 
at contact A have relaxed at the drain reference contact. Therefore, the
chemical potentials of both quasi-particle spin species 
($\mu_\mathrm{qp}^{\uparrow,\downarrow}$) and the Cooper pairs $\mu_\mathrm{cp}$ are 
the same, and the drain contact can be considered as a faithful measure of the
equilibrium chemical potential of the superconductor.

\begin{figure}
\includegraphics[width=\columnwidth]{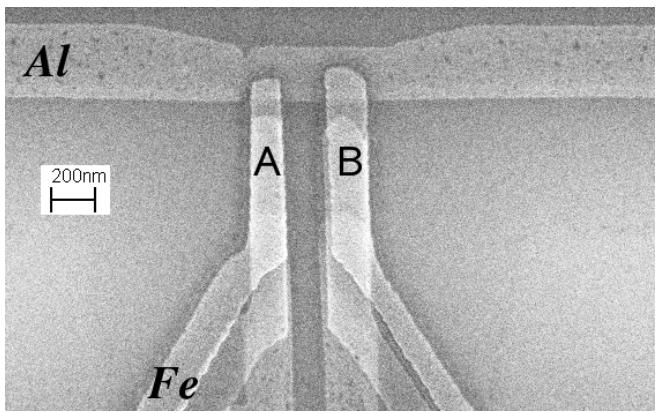}
\caption{\label{fig_SEM}
SEM image of the two contacts of sample II.}
\end{figure}

The experiments were performed on mesoscopic spin-valve structures made 
of ferromagnetic iron and superconducting aluminum by means of e-beam 
lithography and shadow evaporation. The contact region of sample II
is shown in the SEM image in Fig. \ref{fig_SEM}.
In a first evaporation step, 15~nm of iron is evaporated from a direction
almost normal to the plane of an oxidized silicon wafer. The purpose of this first iron
layer is only to ensure Ohmic contacts everywhere in the structure
except for the two tunnel contacts A and B. Then, 25~nm of aluminum are
evaporated under an oblique angle, forming the aluminum wire of 200~nm width
seen in the upper part of the SEM image. The aluminum is then oxidized
in situ admitting a pressure of 60 Pa of pure oxygen for about 1 min. 
Subsequently, a second layer of 20~nm of iron is evaporated from the opposite
direction to form the two tunnel contacts A and B. 
After lift-off, the samples were bonded and mounted into a shielded 
box thermally anchored to the mixing chamber of a dilution refrigerator.
All wire connections from the top of the cryostat to the sample chamber
were filtered through discrete RC low-pass filters, lossy coaxial lines 
and a copper powder filter to eliminate noise and high-frequency thermal
photons from the sample chamber. The experiments presented here were
carried out using a battery and a resistor network as DC current
source, and a nanovoltmeter for voltage measurements. The low-bias results
were confirmed using an AC resistance bridge. Here, we present data from three
samples I, II and III.

\begin{figure}
\includegraphics[width=\columnwidth]{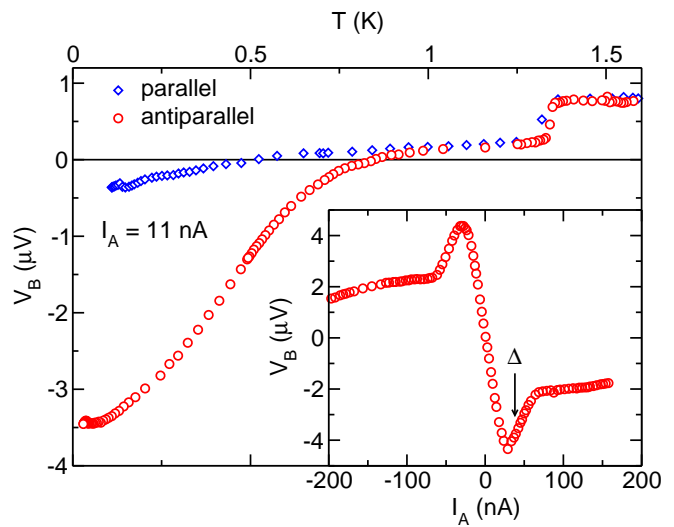}
\caption{\label{fig_T}(Color online) 
Temperature dependence of the detector voltage $V_\mathrm{B}$ at fixed injector current
$I_\mathrm{A}=11~\mathrm{nA}$ for parallel (diamonds) and antiparallel (circles) 
magnetization alignment for sample II. Inset: $V_\mathrm{B}$ as a function of $I_\mathrm{A}$ in the
antiparallel alignment at $T=25~\mathrm{mK}$. $\Delta$ marks the current that corresponds
to $V_\mathrm{A}= 200~\mathrm{\mu V}$, i.e. the gap energy of aluminum.
}
\end{figure}

Figure \ref{fig_T} shows the detector voltage $V_\mathrm{B}$ as a function of temperature
for sample II.
Here, the magnetization of injector and detector contact have been aligned 
parallel or antiparallel at $T>T_\mathrm{c}$ by monitoring the normal-state
spin-valve signal, and then subsequently the sample is cooled down at zero magnetic
field. Above $T_\mathrm{c}$, the observed voltage is essentially due to the
Ohmic resistance of the 10~$\mathrm{\mu m}$ long aluminum strip between contact B and
the drain reference contact, plus the small spin accumulation signal. At 
$T_\mathrm{c}\approx 1.35~\mathrm{K}$, the Ohmic resistance disappears, and
a small positive signal remains around 1~K. At lower temperature, the detector
voltage becomes negative for both magnetization alignments, but with a much larger
magnitude for antiparallel alignment. 

The dependence of the detector voltage $V_\mathrm{B}$ on bias current $I_\mathrm{A}$ is shown
in the inset of Fig. \ref{fig_T}. At low bias current $|I_\mathrm{A}|<30~\mathrm{nA}$, the slope is 
negative, turning into a positive slope at higher bias. By
comparison with the local current-voltage relation (IV) of the injector contact we
see that the slope reversal corresponds to $V_\mathrm{A}\approx200~\mu\mathrm{V}$,
i.e. the superconducting energy gap of aluminum. Thus, we
conclude that the positive slope corresponds to the onset
of the transmission of quasiparticles through allowed
states above the gap.

From the measured non-local voltage as a function of current, the non-local conductance
can be calculated by using the conductance matrix
\begin{equation}
\label{equ_conductance}
\left(\begin{array}{c} I_\mathrm{A} \\ I_\mathrm{B}\end{array}\right) 
=
\left(\begin{array}{cc} G_\mathrm{A} & G_\mathrm{X} \\ 
G_\mathrm{X} & G_\mathrm{B}
\end{array}\right)
\left(\begin{array}{c} V_\mathrm{A} \\ V_\mathrm{B}\end{array}\right),
\end{equation}
which contains the local conductances $G_\mathrm{A,B}$ of the injector (A) and
detector (B) contact, and the non-local conductance $G_\mathrm{X}$. For voltage detection
at B, we have $I_\mathrm{B}=0$, and the current due to the non-local conductance,
$I_\mathrm{X}=G_\mathrm{X}V_\mathrm{A}$, is canceled by the backflow due to the local
conductance, $G_\mathrm{B}V_\mathrm{B}$. By balancing the total currents, we assume that 
all charge carriers emitted into contact B by non-local processes relax to the equilibrium
chemical potential of wire B before they tunnel back. In that case, as $V_\mathrm{B}$ never
exceeds a few microvolts, we can replace $G_\mathrm{B}$ by its low-bias value, and
the energy dependence of $V_\mathrm{B}$ is given only by the dependence of $G_\mathrm{X}$
on $V_\mathrm{A}$. With $G_\mathrm{X}\ll G_\mathrm{A}$, we find

\begin{equation}
G_\mathrm{X}=\frac{dI_\mathrm{B}}{dV_\mathrm{A}}=-G_\mathrm{B}\frac{dV_\mathrm{B}}{dV_\mathrm{A}}.\label{equ_X}
\end{equation}

\begin{figure}
\includegraphics[width=\columnwidth]{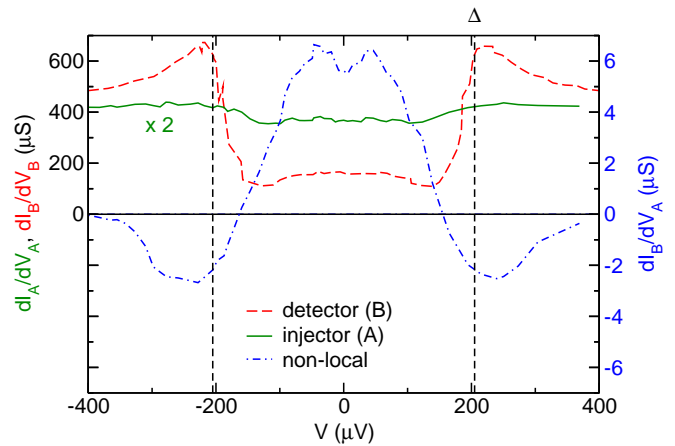}
\caption{\label{fig_dIdV}(Color online) Local differential conductances 
$dI_\mathrm{A}/dV_\mathrm{A}$ 
and $dI_\mathrm{B}/dV_\mathrm{B}$ (left scale), and non-local 
differential conductance $dI_\mathrm{B}/dV_\mathrm{A}$ (right scale), of sample II
at $T \approx 25~\mathrm{mK}$.
$dI_\mathrm{A}/dV_\mathrm{A}$  is doubled for clarity.
}
\end{figure}

The resulting non-local differential conductance $dI_\mathrm{B}/dV_\mathrm{A}$ as
a function of injector voltage $V_\mathrm{A}$, together with the local differential 
conductances of injector and detector contact measured directly, is shown 
in Fig. \ref{fig_dIdV}.

The local conductance of the detector exhibits a suppression at subgap energies,
and peaks at the energy of the gap, $\Delta \approx 200\mu\mathrm{eV}$, which
at least qualitatively corresponds to the expected BCS tunneling characteristics. For
the injector, neither the subgap suppression nor the peaks are well resolved.
For both  injector and detector, the subgap conductance is much too large
to be compatible with a simple BCS behavior, and has a downward curvature
at low bias, which is not expected for either BCS quasiparticle tunneling,
or Andreev reflection. A strong enhancement of Andreev reflection at low
bias in mesoscopic NS tunnel junctions is known to occur due to the constructive
interference of time-reversed electron and hole trajectories on the normal metal
side of the tunnel junction (reflectionless tunneling).\cite{vanwees1992} While 
reflectionless
tunneling exhibits the decrease of conductance with increasing bias in the
subgap regime that we observe, it should not occur in our structures, as 
time-reversal symmetry is broken in the ferromagnetic electrodes. However, it has been
predicted \cite{hekking1994} that an interference enhancement of Andreev
reflection also occurs on the superconducting side of the interface in restricted
geometries. The formalism used to describe this enhancement is the same as used later to
describe crossed Andreev reflection in diffusive metals.\cite{feinberg2003,bignon2004}

The non-local conductance is positive (corresponding to crossed Andreev reflection)
in the subgap regime, and becomes negative (corresponding to
electron transmission) as the injector voltage approaches
the superconducting gap. At the gap, a negative peak is observed, mimicking the
BCS density of states peak seen in the local conductance. The overall magnitude
of the non-local conductance is about a factor of 100 smaller than the local conductance.
Using the model of twodimensional diffusion\cite{hekking1994,bignon2004}, 
we find that a relative reduction of the non-local conductance 
compared to the local Andreev conductance of two orders of magnitude is realistic. However, using the full 
quantitative expression\cite{hekking1994}, the local Andreev conductance
should be of the order of $G_\mathrm{N}^2R_\Box \approx 1~\mathrm{\mu S}$, 
where $G_\mathrm{N}$ is the normal-state tunnel conductance, and $R_\Box$ 
the normal-state sheet resistance of the superconductor. This estimate is much less 
than the conductances actually observed. We suspect that pin-holes in the oxide 
barrier lead to an enhanced Andreev conductance.

\begin{table}
\caption{\label{tab_exp}Characteristic parameters of our samples.
Average transmission probability $t$ of the tunnel contacts,
elastic mean free path $l_\mathrm{el}$, 
coherence length $\xi$,
normalized contact distance $d/\xi$, 
and inverse diffusion time $\hbar/\tau_\mathrm{D}$.
}
\begin{ruledtabular}
\begin{tabular}{lccccc}
sample & $t$              & $l_\mathrm{el}$ & $\xi$           & $d/\xi$ & $\hbar/\tau_\mathrm{D}$ \\ 
       &                  & $(\mathrm{nm})$ & $(\mathrm{nm})$ &         & $(\mathrm{\mu eV})$ \\ \hline
I      & $8\times10^{-5}$ & 13              & 140             & 1.3     & 120                 \\
II     & $4\times10^{-5}$ & 10              & 120             & 1.7     & 70                  \\
III    & $2\times10^{-5}$ & 5               & 85              & 2.3     & 40                  \\
\end{tabular}
\end{ruledtabular}
\end{table}

\begin{figure}
\includegraphics[width=\columnwidth]{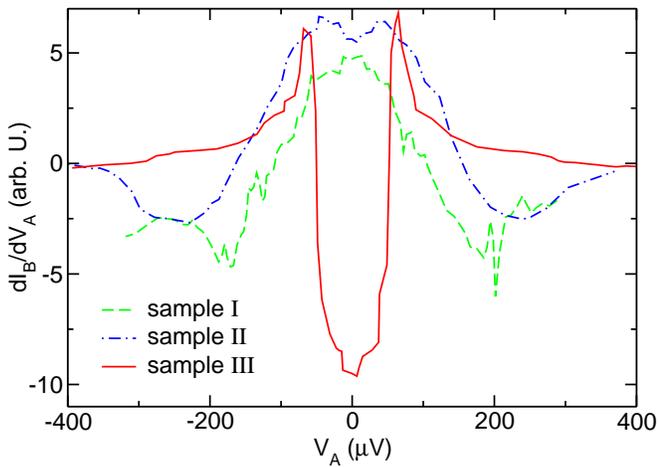}
\caption{\label{fig_X_all}(Color online) Non-local 
differential conductance $dI_\mathrm{B}/dV_\mathrm{A}$ as a function
of injector voltage $V_\mathrm{A}$ for three different samples.
The data are scaled to a similar amplitude.
}
\end{figure}

Figure \ref{fig_X_all} shows the non-local conductance as a function of injector
voltage for three different samples in the antiparallel magnetization state. For 
samples I and II, the data are qualitatively
the same, with a dominating positive signal at low bias, and a negative peak near
the energy gap at about $200~\mathrm{\mu V}$. However, sample III shows a different 
behavior. The non-local conductance is negative at small bias, has a positive peak 
at about $80~\mathrm{\mu V}$, and then decays without any clear feature near the gap.
For sample III, the detector voltage $V_\mathrm{B}$ has the same sign as the injector
voltage over the whole IV trace.

The theoretical predictions for the non-local conductance due to CAR and EC in the
tunneling limit \cite{falci2001} show that both contributions are of the same magnitude and 
opposite sign. Therefore, without spin selection, the non-local
subgap conductance is expected to be zero. Including spin selection a negative
(positive) conductance of equal magnitude is expected for parallel (antiparallel) alignment. 
For increasing contact transparency, EC is predicted to give a larger absolute 
contribution than CAR.\cite{melin2004}
Here, we observe a dominantly positive non-local conductance 
for samples I and II, and  a negative signal for sample III. In Table \ref{tab_exp}, we summarize
the experimental parameters of the three samples. As can be seen, the average contact 
transparency estimated from the normal-state tunnel resistance systematically decreases
from sample I to III. A transition from dominating CAR to dominating EC with decreasing 
transparency is inconsistent with the predictions of Ref. \onlinecite{melin2004}. 
The elastic mean free path $l_\mathrm{el}$ of the aluminum also systematically decreases from 
sample I to III, and 
thereby the coherence length $\xi$ decreases and the normalized contact distance $d/\xi$ increases. The 
dependence of CAR and EC on contact distance is expected to be the same,\cite{falci2001,feinberg2003}
which means that except for an overall signal decrease, no qualitative change is expected with 
increasing $d/\xi$. Recently, it has been predicted \cite{yeyati2006} that
coupling to the electromagnetic environment (dynamical Coulomb blockade) may discriminate
CAR and EC, and also explain a transition between dominating CAR and EC at a finite bias
voltage related to typical energies of environmental modes, either symmetric or antisymmetric with respect
to the two contacts. As the elastic mean free path of the aluminum may affect the 
environmental modes, we speculate that the qualitatively different behavior of sample III compared
to I and II may be related to the decrease in $l_\mathrm{el}$, probably in conjunction with
the decrease in contact transparency.

The non-local conductance of sample III is qualitatively consistent with the observations
made by Russo et al. \cite{russo2005} in a coplanar geometry, where a superconducting
layer was sandwiched between two non-magnetic contacts. Russo et al. report a
positive differential voltage (corresponding to a negative differential conductance) at low bias, 
which reverses sign at an energy scale well below the superconducting gap, and decays
towards higher energies. The energy scale of the sign reversal was found to be the
Thouless energy $E_\mathrm{Th}=\hbar /\tau_\mathrm{D}$, where $\tau_\mathrm{D}$ is the 
diffusion time corresponding to the thickness of the superconducting film (which coincides 
with the distance between the two normal metal contacts in a coplanar geometry). For comparison, 
we have  calculated the energy scale $\hbar/\tau_\mathrm{D}$ corresponding to the diffusion time 
from contact A to B in our samples,
and show the figures in table \ref{tab_exp}. Even though the sign reversal of the non-local conductance
of sample III may be related to $\hbar/\tau_\mathrm{D}$, no clear features are seen at this energy
for sample A and B.

To conclude, we have presented non-local voltage measurements on superconductor-ferromagnet
spin-valve structures, and observed a negative four-probe resistance that provides
unambiguous evidence for crossed Andreev reflection as the dominating subgap 
transport process in some samples. This is an important prerequisite for
the use of crossed Andreev reflection in efficient entanglers. The qualitatively different 
behavior reported by Russo et al. and also observed in one of our samples requires
further systematic experimental investigation as well as theoretical work.

We thank W. Belzig, D. Feinberg, R. M\'elin and A. Levy-Yeyati for useful
discussions, and especially P. Samuelsson, D.
Sanchez, R. Lopez, E. Sukhorukov and M. B\"uttiker for
bringing the source-drain window argument to our
attention. This work was partly supported by the
Deutsche Forschungsgemeinschaft within the Center
for Functional Nanostructures.

\bibliography{../../../lit.bib}

\end{document}